\documentclass[graybox]{svmult}

\usepackage{graphicx}        
\usepackage{multicol}        
\usepackage[bottom]{footmisc}

\usepackage{soul}
\usepackage{newtxtext}       %
\usepackage{newtxmath}       
\usepackage{xcolor}
\usepackage{url} 

\makeindex             

\begin{document}

\title*{Dipolar quantum gases: from 3D to Low dimensions}
\titlerunning{Dipolar quantum gases: from 3D to Low dimensions}
\author{
Yifei He$^1$, Haoting Zhen$^{1,2}$, \and Gyu-Boong Jo$^2$
}

\institute{
$^1$Department of Physics, The Hong Kong University of Science and Technology, Clear Water Bay, Kowloon, Hong Kong China
\and
\\
$^2$Department of Physics and Astronomy,
Smalley-Curl Institute,  Rice University, Houston, TX, USA}
\maketitle

\abstract
{
Dipolar quantum gases, encompassing atoms and molecules with significant dipole moments, exhibit unique long-range and anisotropic dipole-dipole interactions (DDI), distinguishing them from systems dominated by short-range contact interactions. This review explores their behavior across dimensions, focusing on magnetic atoms in quasi-2D in comparison to 3D. In 3D, strong DDI leads to phenomena like anisotropic superfluidity, quantum droplets stabilized by Lee-Huang-Yang corrections, and supersolid states with density modulations. In 2D, we discuss a new scenario where DDI induces angle-dependent Berezinskii-Kosterlitz-Thouless transitions and potential supersolidity, as suggested by recent experimental realizations of strongly dipolar systems in quasi-2D geometries. We identify key challenges for future experimental and theoretical work on strongly dipolar 2D systems. The review concludes by highlighting how these unique 2D dipolar systems could advance fundamental research as well as simulate novel physical phenomena.
}






\newpage

\section{Overview of dipolar atoms and molecules}
\label{sec:1}

Dipolar quantum gases~\cite{lahaye2009physics,Böttcher_2021, chomaz2022dipolar}, such as atoms and molecules with significant dipole moments, have emerged as a vibrant field in ultracold physics. Their long-range and anisotropic dipole-dipole interactions (DDI) clearly distinguish them from systems dominated by short-range contact interactions. 

\subsection{Atoms}
Alkaline atoms typically have magnetic moments around 1$\mu_B$, which contributes to a weak dipolar effect. Dipolar atoms, typically possessing magnetic dipole moments, exhibit anisotropic DDI that influences their quantum gas behavior. The first dipolar Bose-Einstein condensates (BEC) was realized with chromium atoms in 2005~\cite{griesmaier2005bose}. Chromium, with a magnetic moment of 6 $\mu_B$ (Bohr magnetons), provided an early platform for studying dipolar effects. Subsequently, BECs of lanthanide atoms like dysprosium (10 $\mu_B$) and erbium (7 $\mu_B$), which possess larger magnetic moments, were realized, amplifying dipolar effects and enabling richer physics~\cite{lu2011strongly,aikawa2012bose}. More recently, BECs of europium (7 $\mu_B$) were realized~\cite{miyazawa2022bose}.

\vspace{10pt}
{\bf Alkali (weakly dipolar) atoms}
 While alkali atoms (e.g. lithium, rubidium, sodium, potassium) are primarily dominated by contact interactions, their weak magnetic moments can still induce measurable dipolar effects in tailored setups. For example, in spin-polarized alkali BECs, DDI contributes to spin texture formation and collective mode dynamics~\cite{vengalattore2008spontaneously,vengalattore2010periodic,guzman2011long}. While less dominant than in strongly dipolar species, these studies provide insights into the interplay between short- and long-range interactions~
 \cite{Chen.2021a}.

\vspace{10pt}
 {\bf Chromium}
 The BECs of chromium are the first quantum degenerate gases with pronounced DDI~\cite{griesmaier2005bose}. Strong DDI arises from the large electronic spins~($S=6$) of chromium in its electronic ground state. The presence of anisotropic DDI was first observed directly in chromium BECs after Time-of-flight~(TOF) expansions~\cite{stuhler2005observation}. 

\vspace{10pt}
{\bf Lanthanide}
Lanthanide atoms, such as dysprosium and erbium, are cornerstone systems for studying strong DDI due to their large magnetic moments arising from large orbital angular momentum of electrons in their electronic ground state. Although the magnetic moment of chromium is comparable to dysprosium and erbium, the dipolar lengths $a_{dd}=\mu_0\mu^2m/12\pi\hbar^2$ of magnetic lanthanide atoms are much larger than chromium due to the large mass. Dysprosium and Erbium~\cite{lu2011strongly,aikawa2012bose,maier2015broad,ulitzsch2017bose,lucioni2018dysprosium,chalopin2018anisotropic,lunden2020enhancing,seo2023apparatus,song2023dipolar} exhibit pronounced dipolar effects, such as anisotropic superfluidity and the formation of quantum droplets stabilized by beyond-mean-field effects as well as supersolidity. The complex electronic structure of lanthanides also introduces challenges, such as dense Feshbach resonance spectra~\cite{frisch2014quantum,patscheider2022determination,Yijun2015}, necessitating precise control over experimental parameters. 

\paragraph{\bf Two-body interaction potential of polarized dipolar atoms} The total interaction between two polarized magnetic atoms is the sum of contact interaction and DDI, which can be written as

\begin{equation}
    V(\textbf{r})= g^{3D}\delta(r)+\frac{3g_d^{3D}}{4\pi}\dfrac{1-3\cos^2\alpha}{r^3}, 
\end{equation}
where $g^{3D}=\frac{4\pi\hbar^2}{m}a_s$ is the contact coupling constant and $g^{3D}_d=\frac{4\pi\hbar^2}{m}a_{dd}$ is the dipolar coupling constant. $a_s$ is the 3D s-wave scattering length, which can be tuned through the Feshbach resonance. $\alpha$ is the angle between relative position vector $\textbf{r}$ and the polarization axis.

However, we note that representing the two-body interaction as a simple sum of the contact pseudo-potential and DDI potential is actually nontrivial and has been extensively debated~\cite{lahaye2009physics,pitaevskii2016bose}. This expression results from the first-order Born approximation~\cite{yi2000trapped,yi2001trapped}, and is generally valid only for weak dipoles and away from scattering resonances. In the case of strong dipolar interactions, where the relative dipolar strength $\varepsilon_{\text{dd}}=a_{dd}/a_s$ is large, a temperature-dependent correction to $a_{dd}$ may be necessary~\cite{oldziejewski2016properties}.

\subsection{Ultracold polar molecules}
Ultracold polar molecules (KRb, NaRb,...), characterized by large electric dipole moments, offer a complementary platform to dipolar atoms due to their tunable and stronger DDI. Unlike magnetic dipoles, electric dipoles can be induced or controlled via external electric or microwave fields, enabling precise manipulation of the dipolar interaction strength. These molecules, with electric dipole moments up to several Debye, exhibit DDI that can reach a strongly interacting regime with $na_{dd}^3>1$, enabling the realization of strongly correlated states like dipolar crystal or exotic structures~\cite{buchler2007strongly,hertkorn2021pattern}. 

Ultracold polar molecules typically suffer from severe loss, regardless of whether they are chemically reactive~\cite{ospelkaus2010quantum,ni2010dipolar} or not\cite{ye2018collisions,guo2018dipolar,gregory2019sticky}, which causes additional difficulty in cooling and realizing exotic many-body states. Recent advances on electric field shielding~\cite{matsuda2020resonant} and microwave shielding~\cite{karman2018microwave,anderegg2021observation} dramatically increase the elastic collision rate between dipolar molecules, allowing evaporation cooling to the quantum degeneracy for fermionic~\cite{valtolina2020dipolar,schindewolf2022evaporation} and bosonic molecules~\cite{bigagli2024observation}. With stable dipolar molecules in the quantum degenerate regime, various many-body phenomena have been studied such as itinerant spin dynamics~\cite{li2023tunable} in 2D degenerate Fermi gases and strongly interacting self-bound droplets in Bose gases~\cite{zhang2025observation}.

\section{Dipolar gases in a 3D trap}
\label{sec:2}
\subsection{Dipolar collapse}

Anisotropic DDI creates stable regimes that depend on the trap geometry. In a pancake geometry, dipoles along the tight confinement repel each other, which can stabilize the system even with $a_s<0$~\cite{koch2008stabilization}. In contrast, in a cigar trap where dipoles align along the axial direction, attractive DDI makes the system more prone to collapse.

When $a_s$ is tuned below a critical value in a 3D nearly spherical harmonic trap, the mean-field energy becomes negative and peak density increases dramatically until three-body loss dominates. This condition leads to a collapse and dynamic explosion of the gas~\cite{donley2001dynamics}. In chromium BECs, cloverleaf patterns have been observed in TOF images after explosion, reflecting the d-wave shape of the dipolar interaction.

\subsection{Dipolar Quantum droplet}

\begin{figure}[t]
\includegraphics[scale=.2]{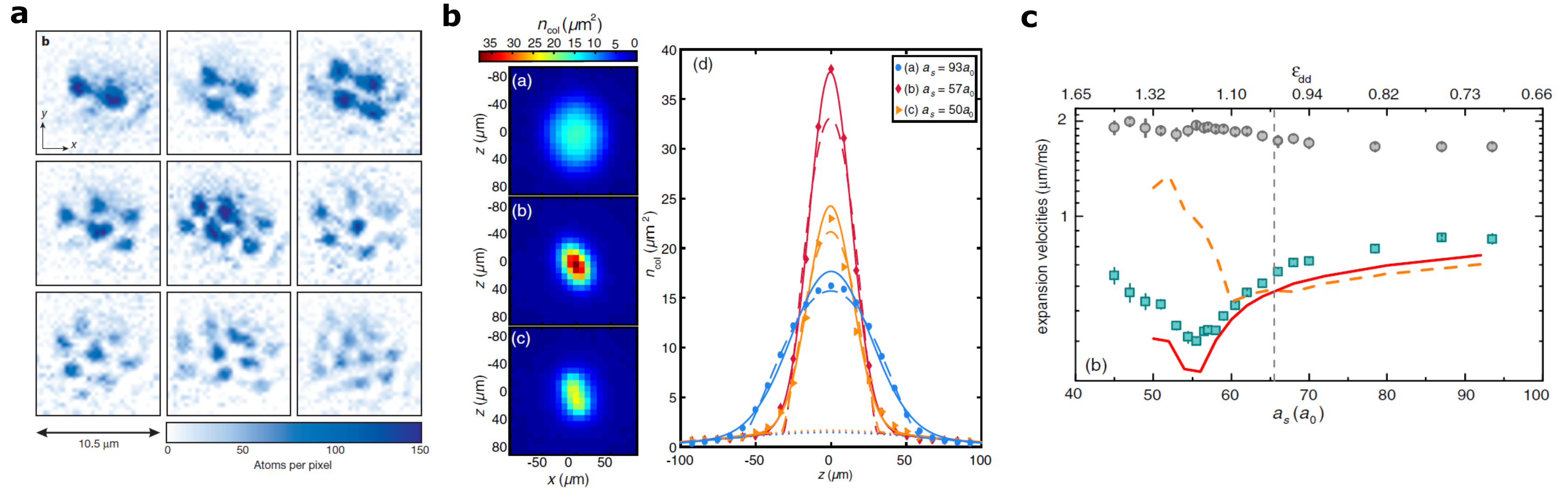}
\caption{The dipolar quantum droplet. \textbf{a} The droplet arrays formed by a pancake-trapped Dy gas with dipoles perpendicular to the tight axis, adapted from \cite{kadau2016observing}. \textbf{b} The single macrodroplet formed by aligning the magnetic B field along the elongated trap. Around the phase transition point between the BEC and quantum droplet phase, the atom density demonstrates a drastic increase (red line). \textbf{c} The expansion velocity of a macrodroplet formed in an cigar-shaped trap. The dip around the BEC-droplet transition point reflects the self-bound nature of the quantum droplet. \textbf{b} and \textbf{c} are adapted from \cite{chomaz2016quantum}.}
\label{droplet_fig}       
\end{figure}

For dysprosium (Dy) and erbium (Er), which exhibit strong DDI, the behavior of BECs deviates significantly from conventional expectations in certain parameter regimes. When the combined mean-field interaction (comprising both contact interactions and DDI) becomes negative, one would typically anticipate the collapse of the dipolar BEC. However, instead of collapsing, the system forms a stable, self-bound quantum droplet~\cite{schmitt2016self,chomaz2016quantum} or even organized droplet arrays~\cite{kadau2016observing,schmitt2016self}. This unexpected stability arises from the beyond-mean-field quantum fluctuations, specifically the Lee-Huang-Yang (LHY) correction, which plays a critical role in stabilizing the system~\cite{lee1957eigenvalues,lee1957many}.  A self-bound droplet with a similar stabilization mechanism is also observed in short-range interacting mixtures (see chapter 8). Under normal conditions, the LHY correction is relatively small and often negligible. However, in erbium (Er) and dysprosium (Dy) systems, when the mean-field contact interaction and DDI nearly cancel each other out (typically when $a_s$ is comparable to $a_{dd}$), the LHY term becomes dominant. This dominance prevents the system from undergoing severe three-body losses, thereby stabilizing the droplet. The interplay between the mean-field interactions and the LHY correction allows these droplets to maintain their structural integrity, exhibiting unique properties, which have been experimentally observed, as shown in Fig.\ref{droplet_fig}. These quantum droplets represent a fascinating state of matter, bridging the gap between mean-field dynamics and quantum fluctuations.

\subsection{Elementary excitations}
\begin{figure}[b]
\includegraphics[scale=.60]{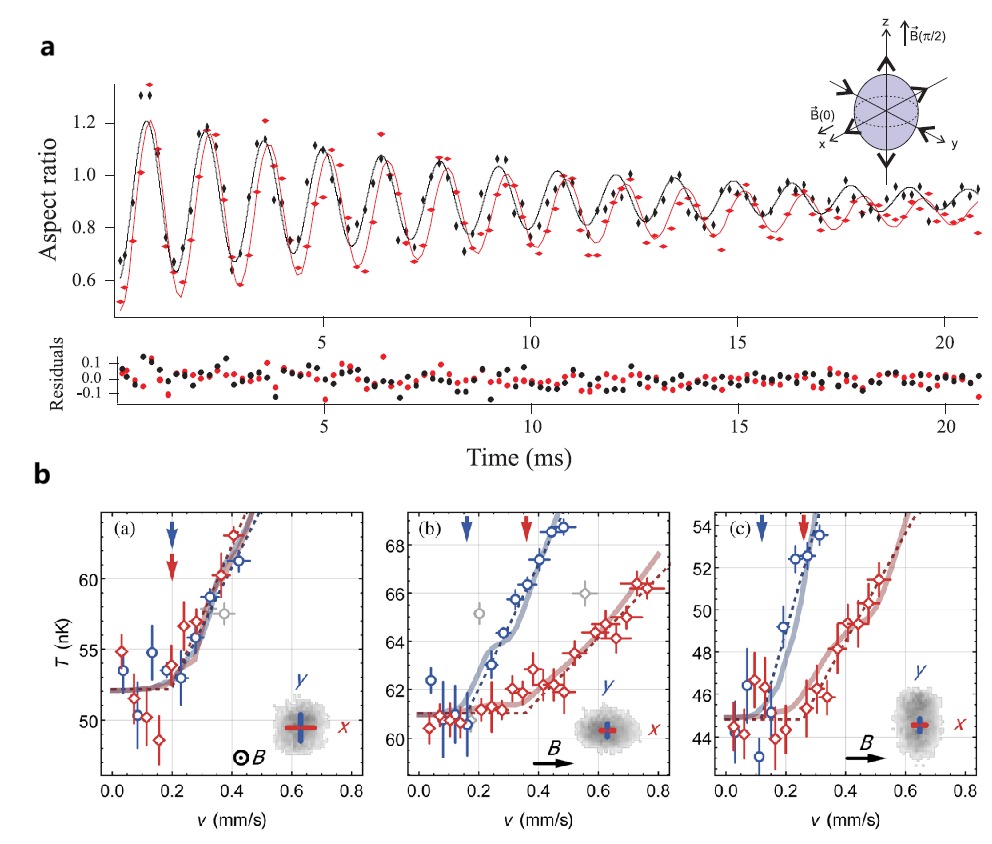}
\caption{The anisotropic elementary excitations of dipolar gas. \textbf{a} The discrepancy of quadrupole mode frequency of a chromium gas with two perpendicular dipole orientations, adapted from \cite{bismut2010collective}. The inset shows the illustration of the 3D quadrupole mode, which is observed from the oscillation of the aspect ratio along y and z axis. The magnetic field is either along z axis (corresponding data shown in black) or along x axis (data shown in red). The lower panel shows the residual of the sinusoidal fittings, shown in solid lines. \textbf{b} The superfluid critical velocity in a $^{162}\textrm{Dy}$ BEC becomes anisotropic due to the anisotropic DDI, adapted from \cite{wenzel2018anisotropic}.}
\label{excitation_fig}       
\end{figure}

DDI also gives rise to novel many-body effects in dipolar BECs, for example, shifting the quadrupole mode of collective excitation~\cite{bismut2010collective}, as shown in Fig \ref{excitation_fig}a. The Fourier transform of eq.(1) gives the interaction in momentum space:

\begin{equation}
    V(\textbf{k})=\frac{4\pi\hbar^2}{m}\left[a_s+a_{dd}(3\cos^2\theta_k-1)\right],
\end{equation}
where $\theta_k$ is the angle between the direction of momentum and the dipole orientation, resulting in an elementary excitation spectrum:
\begin{equation}
    E(\textbf{k})=\sqrt{\frac{\hbar^2k^2}{2m}\left(\frac{\hbar^2k^2}{2m}+2nV(\textbf{k})\right)}.
\end{equation}
This anisotropic excitation and corresponding anisotropic superfluidity are observed through Bragg spectroscopy
~\cite{bismut2012anisotropic} and the critical superfluid velocity respectively
~\cite{wenzel2018anisotropic}, shown in Fig \ref{excitation_fig}b.

\subsection{Roton spectrum induced by anisotropic confinement}
\begin{figure}[b]
\includegraphics[scale=.55]{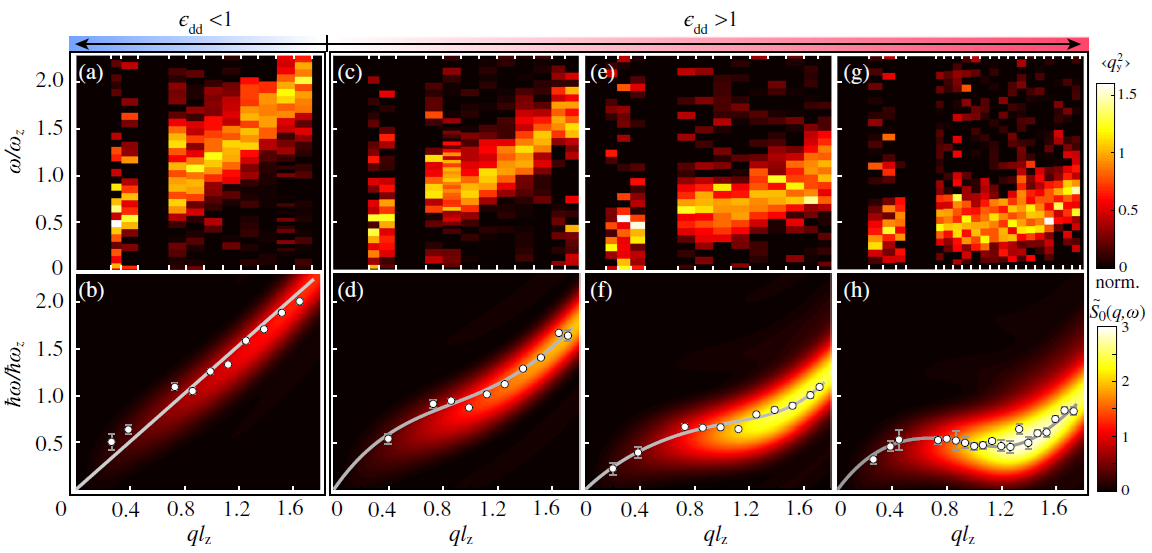}
\caption{The excitation spectrum of an elongated dipolar gas with different $\varepsilon_{\text{dd}}$. The top row shows the experimental data of Bragg spectroscopy, the bottom row presents the extracted excitation spectrum $\varepsilon(q)$ (white dots) and simulated structure factor. Both panels demonstrate the rotonic behavior when $\varepsilon_{\text{dd}}>1$ where the DDI dominates, adapted from \cite{petter2019probing}. The solid lines represents the analytical curve adapted from \cite{2012Blakie}.}
\label{roton_fig}       
\end{figure}

In uniform 3D dipolar BECs, the excitation energy is dependent on the orientation of the relative momentum but independent of the magnitude of the momentum. More striking behaviors of the excitation spectrum arise when anisotropic confinement is introduced along the dipoles. The interplay of the DDI with the harmonic oscillator length of the tightly confined direction of the trap $l_z$ leads to a $k$-norm-dependence along the unconfined direction:
\begin{equation}
    V(k)\sim ng_c\left[1+\varepsilon_{\text{dd}}F(\frac{kl_z}{\sqrt{2}})\right],
\end{equation} 
where $F(x)=2-3\sqrt{\pi}xe^{x^2}\rm erfc(\textit{x})$ is a monotonically decreasing function. It rapidly decreases from $F(0)=2$ to $F(1)\sim0$, and finally it asymptotically approaches $-1$ at $\mathrm{lim}_{x\rightarrow\infty} $. With sufficient large DDI and dipoles along the tight axis, a non-monotonic dispersion relation along the unconfined direction forms, presenting a local maximum~(maxon) and local minimum~(roton) at finite momentum, as was observed in superfluid liquid helium. This effect is due to the fact that DDI~(also seen by $V(\textbf{k})$) becomes attractive when $kl_z\gg1$. The location of the roton minimum scales as the harmonic oscillator length of the tightly confined direction $ k_{\text{rot}}\propto 1/l_z$~\cite{santos2003roton}. The roton gap $\Delta$ can be reduced by increasing $\varepsilon_{\text{dd}}$ or density. When $\Delta=0$, atoms can spontaneously populate in the roton mode $k_{\text{rot}}$ without additional energy cost
~\cite{chomaz2018observation}, thus forming a roton instability. The roton excitation spectrum was observed through Bragg spectrum in Er BECs in an elongated cigar-shaped trap geometry as shown in Fig.~\ref{roton_fig}~\cite{chomaz2018observation,petter2019probing}. As the roton instability is approached~($E(k_{\text{rot}})\approx0$), the density-density correlation is enhanced at $k_{\text{rot}}$, indicating the tendency of the system to crystallize~\cite{hertkorn2021density}. 

\subsection{Supersolid}
\begin{figure}[b]
\includegraphics[scale=.38
]{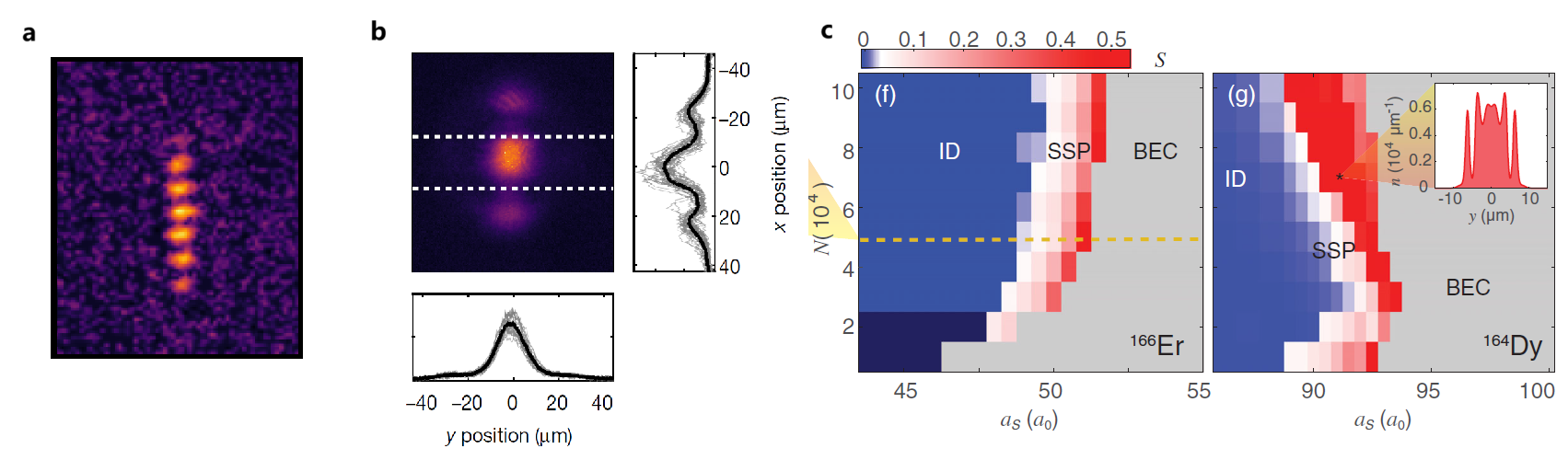}
\caption{The supersolidity emerging in an elongated dipolar gas. \textbf{a.} The \textit{in situ} image of a supersolid in a elongated trap, which has clear density modulation. \textbf{b.} The TOF inteference pattern formed from the supersolid sample in \textbf{a}. After averaging, the uniaxial modulation reflects the phase coherence of the original sample. Both \textbf{a} and \textbf{b} are adapted from \cite{norcia2021two}. \textbf{c.} The phase diagram of Er and Dy gas in an elongated trap. With decreasing $a_s$, the gas transforms from BEC to supersolid state~(SSP), and eventually becomes insulating droplets~(ID), adapted from \cite{chomaz2019long}.}
\label{supersolid_fig}       
\end{figure}

In dipolar BECs, the roton instability leads to the population of the roton mode at $k_{\text{rot}}$, producing density modulations with spatial period $\sim2\pi/k_{\text{rot}}$ on the superfluid background, as shown in Fig.~\ref{supersolid_fig}. This spontaneous breaking of the translational symmetry and $U(1)$ symmetry indicates a long-predicted supersolid state~\cite{tanzi2019observation,bottcher2019transient,chomaz2019long}. The beyond-mean-field LHY term again plays an important role to prevent the density peaks from growing to an infinitely high value. The supersolid phase could also be understood as droplet arrays immersed in the BEC background sharing a global phase through particle exchange, thus simultaneously breaking the U(1) gauge symmetry. Several key features of the supersolidity of the dipolar crystal are confirmed, including global coherence~\cite{bottcher2019transient}, phase stiffness~\cite{guo2019low}, and simultaneous breaking of two symmetries~\cite{tanzi2019supersolid}.  Density fluctuations~\cite{hertkorn2021density,schmidt2021roton} , superfluid fraction \cite{tanzi2021evidence,biagioni2024measurement}
 and quantized vortices in supersolids~\cite{casotti2024observation} were further studied, providing a comprehensive understanding of the dipolar supersolid in 3D. 
 
 By tuning the trap geometry and atom number, supersolids with more complex structures beyond one-dimensional arrays could form. In an oblate trap with dipoles along the axial direction, supersolids with zigzag or hexagonal structures were observed~\cite{norcia2021two,bland2022two}. We notice that those supersolids have two-dimensional structures, but the tight confinement is always of the order of 100 Hz, making the wave functions still 3D. The superfluidity of these supersolid states with 2D structures can still be attributed to the Bose-Einstein condensation driven by quantum statistics, and the interplay between supersolid and 2D superfluid has not been addressed~\cite{hadzibabic2008trapped}.  The formation of roton is essentially dependent on the competition between the positive term $\hbar^2k^2/2m+2ng_c(1+2\varepsilon_{\text{dd}})$, which describes the kinetic energy and the effective short-range interaction, and the negative term $-3ng_{d}\sqrt{\pi}(kl_z)e^{(kl_z)^2}\mathrm {erfc}(kl_z)$, which captures the long-range nature of DDIs in confined geometry, in the excitation energy $E(k)=\sqrt{\frac{\hbar^2k^2}{2m}\left(\frac{\hbar^2k^2}{2m}+2nV({k})\right)}$ with $V(k)$ given by Eq.~(4). The choice of a "fat" oblate or surfboard trap is simply because a relatively large $l_z$ makes $E(\textbf{k}_\text{rot})$ easier to soften to 0 with $k_{\text{rot}}l_z\sim1$. For the same reason, supersolids in the strict quasi-2D regime where $l_z$ is smaller than the healing length $\xi$ usually need an extremely large atom number~(2D density) or dipolar strength~\cite{zhang2021phases,ripley2023two}, which is inaccessible under current experimental conditions and has not been observed to date.

\section{Two-dimensional Dipolar gases}

Ultracold fermionic molecules have been arranged in 2D traps to demonstrate the collisional stable gas with repulsive DDI~\cite{valtolina2020dipolar} and study itinerant spin dynamics~\cite{li2023tunable}. 2D confinement leads to a dipolar shielding effect which suppresses the dipolar relaxation~\cite{barral2023can}. Interlayer coupling between two layers of 2D dipolar gases has been observed in bilayer dipolar systems with 50~nm interlayer spacing~\cite{du2024atomic}.

In parallel, the Berezinskii-Kosterlitz-Thouless (BKT) transition, a hallmark of 2D superfluidity, has been investigated in systems with strong DDI. The BKT transition was first observed in 2D dipolar excitons, where the dipoles are oriented perpendicular to the plane of confinement~\cite{dang2019defect,dang2020observation}. More recently, 2D dipolar superfluids with fully controllable dipole orientations have been realized using ultracold erbium atoms~\cite{he2025exploring}. The observation confirms the prediction of monte-carlo calculations~\cite{filinov2010berezinskii,bombin2019berezinskii}. These findings not only validate theoretical models but also pave the way for exploring novel quantum phases, such as anisotropic superfluids, exotic vortex dynamics and possible supersolids, in ultracold dipolar systems.

\subsection{Interaction in a 2D dipolar gas}
\subsubsection{Interaction in a pure-2D dipolar gas}
Before examining dipolar atoms trapped in a quasi-2D trap, it's instructive to consider dipoles with mass $m$ moving in the $x-y$ plane that interact only through DDI. An external magnetic (or electric) field in the $y-z$ plane polarizes all dipoles in the same direction with angle $\theta$ with respect to the $z$ axis. The two-body interaction between dipoles $i,j$ can be written as
\begin{equation}
V(\textbf{r})=\frac{C_{dd}}{4\pi}\frac{1-3\sin^2\theta\cos^2\phi_{ij}}{r^3_{ij}},
\end{equation}
where ($r_{ij},\phi_{ij}$) is the polar coordinate of $\textbf{r}_{ij}$, $C_{dd}=d^2/\epsilon_0~(\mu_0\mu^2)$ for electronic (magnetic) dipoles. The characteristic dipolar length is defined as $r_d=mC_{dd}/(4\pi\hbar^2)$. With only DDI, there is a critical angle $\theta_c=35.3^\circ$ above which the system becomes unstable. Because when $\theta_c>35.3^\circ$, two dipoles approaching each other with $\phi_{ij}=0^\circ$ have attractive interactions, which leads to a bound state in pure-2D geometry.

Whether an interaction is long-range is dependent on the problems we consider as well as on the dimensionality~\cite{chomaz2022dipolar}. In pure 2D geometries, the DDI is marginally short-ranged when the dipolar strength is weak, and the two-body scattering between dipoles can be well approximated by s-wave scattering with 2D scattering length $a_s=\frac{e^{2\gamma}}{2}(3\cos^2\theta-1)r_{d}$~\cite{macia2011microscopic}, where $\gamma\approx0.57721...$ is Euler's gamma constant. In the dilute regime where $na_s^2\ll1$, an effective hard disk model adequately captures the system properties at the many-body level, preserving the universal scaling properties of the energy per particle~\cite{macia2011microscopic}. When the system enters the strongly interacting regime ($nr_d^2>1$), the long-ranged nature of DDI begins to manifest itself as the roton spectrum begins to form and crystallization tendencies emerge~\cite{macia2012excitations}.

\subsubsection{Interaction in a Quasi-2D dipolar gas}

The quasi-2D regime is more commonly used in ultracold atom experiments than the pure-2D regime. When a dipolar gas meets the quasi-2D condition—where the wave function along $z$ is frozen into a Gaussian distribution with oscillator length $l_z=\sqrt{\hbar/m\omega_z}$—the inter-particle interaction potential can still be treated in 3D form because $a_{dd},a_s\ll l_z$.

\begin{figure}[b]
  \centering
\includegraphics[scale=0.35]{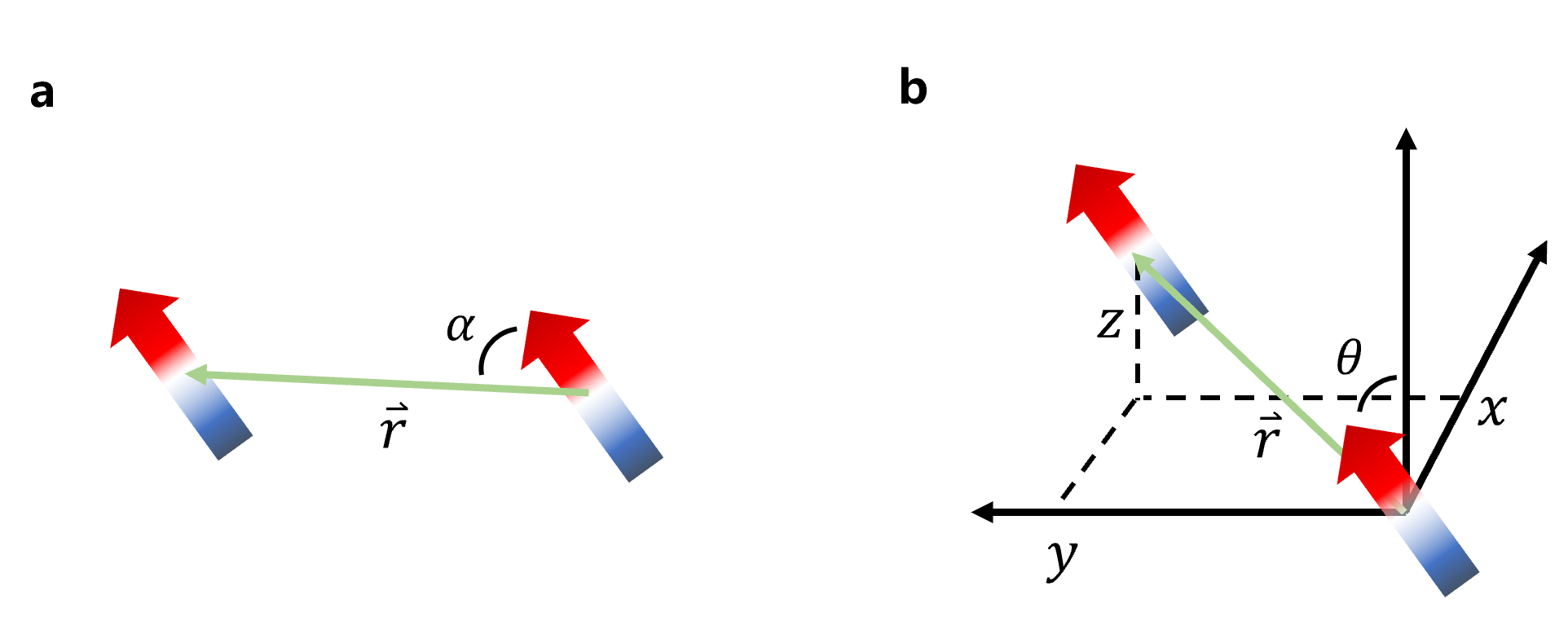}
\caption{\textbf{a.} For two polarized dipoles, DDI is determined by the angle $\alpha$ between dipolar orientation and relative position $\textbf{r}$, as desribed by Eq.~(1). \textbf{b.} Considering two dipoles are polarized by angle $\theta$ with respect to the $z$ axis in the $y-z$ plane, it holds that $\cos^2\alpha=z^2\cos^2\theta/r^2+y^2\sin^2\theta/r^2$.}
\label{DDI}       
\end{figure}

Consider all the dipoles are polarized to angle $\theta$ with respect to the $z$ axis and tilted toward the $y$ axis, then Eq.~(1) can be written as:
\begin{equation}
    V(\textbf{r})= g^{3D}\delta(r)+\frac{3g_d^{3D}}{4\pi}\dfrac{1-3z^2\cos^2(\theta)/r^2-3y^2\sin^2(\theta)/r^2}{r^3}, 
\end{equation}
where $g_d^{3D}=\frac{4\pi\hbar^2}{m}a_{dd}$ and $\theta$ is the angle between the dipoles and the $z$ axis in the $x-y$ plane, $\textbf{r}=(x,y,z)$ is the relative position vector, $r=\lvert{\textbf{r}}\rvert$, as shown in Fig.~\ref{DDI}.  After a Fourier transform,
\begin{equation}
    V(\textbf{q})=\int d^3x e^{i\textbf{q}\cdot\textbf{r}}V(\textbf{r}),
\end{equation}
we have
\begin{equation}
    V(\textbf{q})=g^{3D}+g_d^{3D}\left[\dfrac{3q_z^2\cos^2(\theta)+3q_y^2\sin^2(\theta)}{q_x^2+q_y^2+q_z^2}-1\right].
\end{equation}
In the above expression, $\textbf{q}$ is in 3D. In the experiment, the system is tightly confined in the $z$ direction through a harmonic trap, therefore it is convenient to convert one label of $V(q)$ into the harmonic label $n$ and $n'$. We define
\begin{equation}
    \tilde{V}(\textbf{k};n,n')\equiv \int \frac{dq_z}{(2\pi)}|\phi_n|^{2}|\phi_{n'}|^{2}V(\textbf{q}),
\end{equation}
where $\textbf{k}$ signifies that the vector only has two dimensions. When quasi-2D condition is fulfilled, we take single mode approximation that only considers the interaction in the ground state.
In momentum space representation,
\begin{equation}
    |\phi_0|^2(q_z)=\exp\left(-\frac{1}{4}q_z^2l_z^2\right),
\end{equation}
thus
\begin{equation}
\begin{aligned}
    \tilde{V}(\textbf{k};0,0)&=\int\frac{dq_z}{(2\pi)}\exp\left(-\frac{q_z^2l_z^2}{2}\right)V(\textbf{q})\\
    &=\frac{(g^{3D}-g_d^{3D}+3\cos^2\theta g_d^{3D})}{\sqrt{2\pi}l_z}+\frac{3}{2}g_d^{3D}ke^{k^2l_z^2/2}\mathrm{erfc}(kl_z/\sqrt{2})\left[\frac{k_y^2\sin^2\theta}{k^2}-\cos^2\theta\right].
\end{aligned}
\end{equation}
With redefined contact 2D coupling constant $g_c=\sqrt{8\pi}\hbar^2a_sm/l_z$ and dipolar 2D coupling constant $g_d=\sqrt{8\pi}\hbar^2a_{dd}/ml_z$, we have
\begin{equation}
    V(\textbf{k})=g_{\text{eff}}+3g_dG(kl_z/\sqrt{2})\left[\frac{k_y^2\sin^2\theta}{k^2}-\cos^2\theta\right],
\end{equation}
where $G(q)\equiv \sqrt{\pi}qe^{q^2} {\rm erfc} (q)$, $g_{\text{eff}}=g_c+g_d(3\cos^2\theta-1)$, and $\rm erfc$ representing the complementary error function. The function $G(q)$ is zero when $q=0$ and asymptotically approaches 1 when $q\gg1$.

From $V(\textbf{k})$ we can see that the total interaction in 2D is decomposed into an isotropic local part and an anisotropic nonlocal part. The local part $g_{\text{eff}}=g_c+g_d(3\cos^2\theta-1)$ includes the contribution from both contact interaction and the short-range part of DDI, which contains an effective contact interaction contribution. The momentum-dependent term $3g_dG(kl_z/\sqrt{2})\left[k_y^2\sin^2\theta/{k^2}-\cos^2\theta\right]$ is the long-range part of the DDI, which vanishes at $\textbf{k}=0$. Specifically, $-3g_dG(kl_z/\sqrt{2})\cos^2\theta$ has an attractive nature at large momentum, which gives the tendency to form a roton spectrum when dipoles are aligned out of the plane~($\theta<90^\circ$). In contrast, $3g_dG(kl_z/\sqrt{2})k_y^2\sin^2\theta/{k^2}$ is repulsive, leading to an anisotropic excitation spectrum when dipoles are tilted ($\theta>0^\circ$). In a zero-temperature homogeneous 2D dipolar condensate, the system recovers a mean-field solution $\mu=g_{\text{eff}}n$ as all the particles populate the mode $\textbf{k}=0$. With only DDI, the stable boundary of homogeneous 2D dipolar condensate is given by $g_d(3\cos^2\theta-1)>0$, giving a critical angle $\theta_c=54.7^\circ$ which is different from the pure-2D condition~\cite{mishra2016dipolar}. When density-inhomogeneity is induced by an external harmonic trap or structures such as vortices and solitons, the non-local term of DDI could manifest itself~\cite{cai2010mean,mulkerin2013anisotropic,chen2021observation}.

Generally speaking, the short-range nature of DDI becomes more dominant when $l_z$ decreases. Considering the crossover between quasi-2D and pure-2D regime in the limit of $l_z\rightarrow0$, the mechanism of roton formation changes. Specifically, the source of roton spectrum in quasi-2D is the attractive interaction at short distances, while in pure-2D is the short-range order led by the strong repulsive interaction~\cite{hufnagl2011roton}.

\subsection{Berezinskii-Kosterlitz-Thouless~(BKT) transition in a 2D dipolar Bose gas}

 The BKT transition is an interaction-driven infinite order superfluid transition in 2D systems that occurs without continuous symmetry breaking at finite temperature. This distinguishes it from the statistics-driven BEC transition in 3D, which breaks a U(1) gauge symmetry. In 2D contact-interacting gases, the BKT transition has been experimentally observed and studied~\cite{hadzibabic2006berezinskii,hung2011observation,murthy2015observation,fletcher2015connecting,sunami2022observation}.

A more intriguing question concerns how superfluidity emerges in a 2D dipolar gas at finite temperature~\cite{filinov2010berezinskii}. These systems feature long-range dipolar interactions that induce anisotropy~\cite{bombin2019berezinskii}. Various theoretical models have been developed to study 2D dipolar gases, examining aspects such as stability~\cite{Fischer.2006} and universality~\cite{Vasiliev.2014}. The Hartree-Fock mean-field method and the zero-temperature extended Gross-Pitaevskii equation (eGPE) have been often employed to investigate both thermal and deep superfluid regimes. However, studying the BKT critical point accurately at finite temperature requires incorporating beyond mean-field correlations~\cite{holzmann2008kosterlitz,holzmann2010universal}.

 \begin{figure}[b]
 \centering
\includegraphics[scale=0.33]{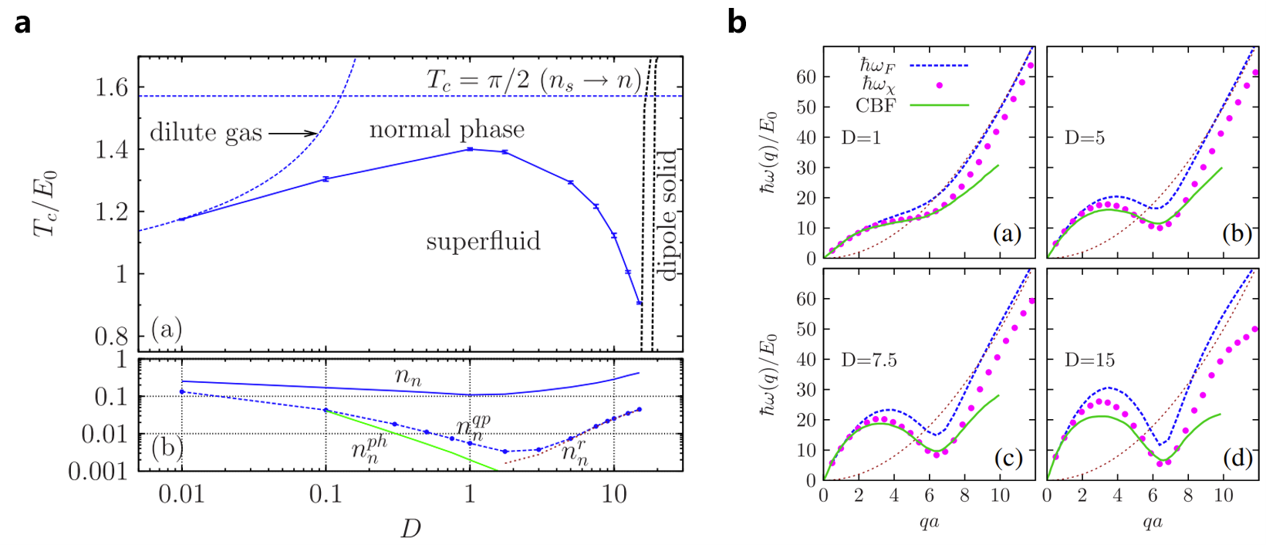}
\caption{\textbf{a.} Nonmonotonic behavior of BKT critical temperature (points connected with solid lines) as a function of dipolar strength. $D=nr_d^2$, calculated by PIMC. \textbf{b.} Excitation spectrum with different dipolar strength $D$. Roton spectrum appears in the strongly interacting regime, and the onset of roton minimum coincides the turning point of critical temperature in Fig.~\ref{MC_BKT}a. Adapted from Ref.~\cite{filinov2010berezinskii}.}
\label{MC_BKT}       
\end{figure}

\subsubsection{Finite temperature simulation of (quasi-)2D dipolar gases}

An effective method to deal with the finite temperature system is the Monte Carlo simulation. BKT transition is predicted to occur with DDI by path integral Monte Carlo simulation~(PIMC) in the pure-2D regime~\cite{filinov2010berezinskii,bombin2019berezinskii}. DDI can be considered as a short-range interaction in the dilute regime $nr_d^2\ll1$, and the BKT critical point can be reproduced by an s-wave scattering length with asymptotic expression $a_s=\frac{e^{2\gamma}}{2}(3\cos^2\theta-1)r_{d}$~\cite{filinov2010berezinskii,bombin2019berezinskii}. When the interaction strength increases, the superfluid phase is first stabilized, and then tends to form a dipolar crystal where the superfluid density vanishes, as shown in Fig.~\ref{MC_BKT}a. The critical temperature of the superfluid transition shows a nonmonotonic behavior. Fig.~\ref{MC_BKT}b shows the excitation spectrum with different dipolar strength, where then roton-maxon spectrum emerges when $D>1$. The maximum transition temperature in Fig.~\ref{MC_BKT}a is reached at $D=1$, coinciding with the appearance of a roton minimum in the excitation spectrum.

The Hartree-Fock-Bogoliubov method with the Popov (HFBP) approximation~\cite{ticknor2012finite,ticknor2012anisotropic} offers another approach that incorporates both finite temperature effects and beyond mean-field effects in quasi-2D dipolar gases. This HFBP reveals anisotropic phase-phase correlations and compressibility in 2D dipolar Bose gases when dipoles are tilted. Despite these advances, no theoretical prediction of the BKT transition point has been made in a quasi-2D dipolar gas.

\subsubsection{Dipolar BKT superfluid}

\begin{figure}[b]
\includegraphics[scale=0.33]{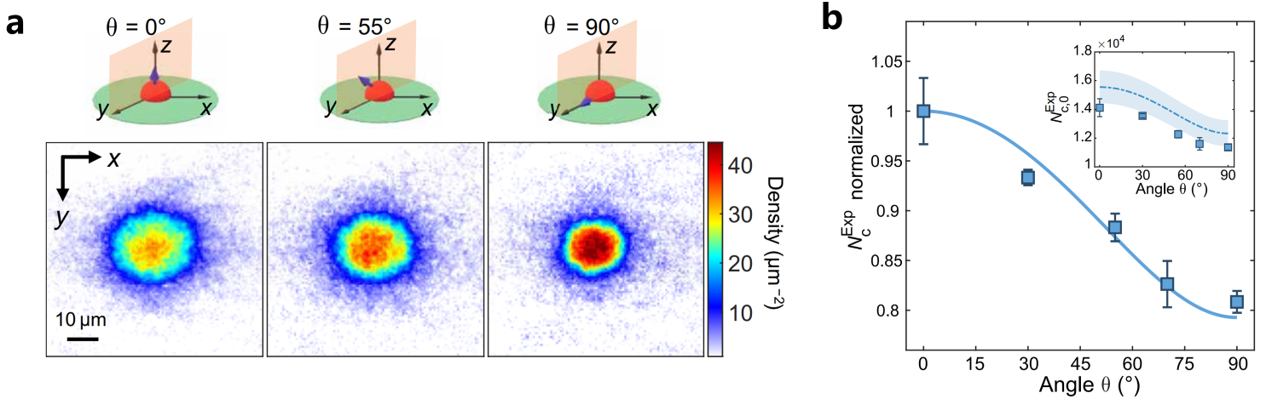}
\caption{\textbf{a.} \textit{In situ} images of 2D dipolar superfluid of erbium atoms with $\varepsilon_{\text{dd}}=0.47$. \textbf{b.} Relative shift of critical atom number of BKT transition as a function of dipolar orientation $\theta$. The solid curve is the critical temperature predicted for local coupling $\tilde{g}_{\text{eff}}(\theta)$ as a function of $\theta$~\cite{holzmann2008kosterlitz}. Inset is the absolute critical atom number measured in experiments. Adapted from Ref.~\cite{he2025exploring}.}
\label{BKT_Tc}       
\end{figure}

In the quasi-2D regime, the BKT transition of the weakly interacting dipolar gas with both contact interaction and DDI has been observed~\cite{he2025exploring}. With dominant local interaction at $\varepsilon_{\text{dd}}\sim0.5$, it was demonstrated that DDI is involved in the BKT transition by acting as an effective contact interaction. When $\theta$ is tuned from $0^\circ$ to $90^\circ$, the atomic clouds shrink almost isotropically, consistent with the decrease in isotropic ${g}_{\text{eff}}$ defined in Eq.~(12)~(Fig.~\ref{BKT_Tc}a). By monitoring the onset of coherence in momentum space, it was shown that the dipolar angle $\theta$ shifts the superfluid transition point, which can be approximately described by the BKT transition point predicted for local coupling ${g}$, with ${g}_{\text{eff}}(\theta)$ as a function of $\theta$~(Fig.~\ref{BKT_Tc}b). This observation of an angle-dependent transition point aligns with PIMC calculations~\cite{bombin2019berezinskii}. In the dipolar BKT superfluid, anisotropic atom number fluctuations emerge when dipoles are predominantly tilted into the 2D plane, signaling an anisotropic density-density correlation~\cite{he2025exploring,baillie2014number}.

\subsection{Strongly dipolar regime in 2D}

When dipolar interactions become dominant (e.g. $\epsilon_{dd} \simeq 1$), it remains unclear how they modify the intrinsic properties of 2D contact-interacting gases. These properties include scale invariance with its relevant collective excitations and energy spectrum with instabilities. Additionally, whether the BKT scenario remains valid for superfluidity in this strongly dipolar 2D case is uncertain. The same holds for vortex-pairing dynamics and topological phase transitions under these conditions. Particularly interesting scenarios emerge when the dipoles are tilted, potentially leading to the emergence of a supersolid phase.

\subsubsection{Scale invariance }
Scale invariance is a macroscopic property of 2D quantum gases, resulting from the length-scale-agnostic classical-field description of low-energy atomic scattering~\cite{pitaevskii1996dynamics}. This scale-invariant behavior has been observed in 2D BKT superfluids~\cite{hung2011observation, saint2019dynamical}. However, it breaks down when interaction details must be considered, as seen in strongly interacting 2D Fermi gases~\cite{peppler2018quantum,holten2018anomalous}. 

In a 2D dipolar gas, the nonlocality of DDI introduces an extra \textbf{k}-dependent term even in the weakly-interacting regime, as shown in Eq.~(11). Therefore, the scale invariance should not hold in a 2D gas where DDI is significant. Although the experimental studies on 2D BKT superfluid and 2D Townes solitons suggest minimal deviation from scale-invariant behavior
~\cite{he2025exploring,chen2021observation}, the 2D dipolar gas with stronger DDI ($\varepsilon_{\text{dd}}\gtrsim 1$) is expected to violate such a scaling symmetry, and more generally, a conformal SO(2,1) symmetry in a harmonic trap
~\cite{pitaevskii1997breathing}. For contact interacting gases in 2D harmonic trap, . This DDI-induced property may be utilized in the simulation of early universe, as proposed in the works of Refs.
~\cite{cha2017probing, chandran2025expansion}.  Collective excitations would serve as versatile tools for characterizing strongly dipolar 2D systems similar to contact-interacting gases in 2D~\cite{peppler2018quantum,holten2018anomalous,He.2020}. Their behavior—whether conforming to scale invariance or deviating from it—is therefore critical to understanding these systems. Recently, the breaking of scale invariance was observed in a 2D strongly dipolar superfluid of erbium with tilted dipoles by measuring the monopole mode frequency~\cite{zhen2025breaking}. 

\subsubsection{Instabilities}
When ${g}_\text{eff}=g_c+g_d(3\cos^2\theta-1)$ is a sufficiently positive value in a dilute 2D dipolar condensate, the system stays in a stable superfluid regime, especially in the case where the dipoles are perpendicular to the 2D plane
~\cite{fischer2006stability}. One can evaluate how instabilities occur in homogeneous 2D dipolar condensate at mean-field level. Apparently, one can make $g_{\text{eff}}$ close to zero or even below zero by either decreasing $g_c$ to negative while keeping the dipoles perpendicular to the 2D plane, or tilting the dipoles in the strongly dipolar regime with $\varepsilon_{\text{dd}}\gtrsim1$. When $g_{\text{eff}}$ is positive but close to zero, the system could enter a roton instability regime with $\theta<90^\circ$ due to the attractive non-local tail $-3ng_{d}\sqrt{\pi}(kl_z)e^{(kl_z)^2}\mathrm {erfc}(kl_z)\cos^2\theta$ coming from DDI~\cite{mishra2016dipolar}. When $g_{\text{eff}}<0$, the system enters a phonon instability regime~\cite{mishra2016dipolar}.
When tilting the dipoles to let the system enter the roton instability regime, the anisotropic excitation spectrum will only cause a roton minimum along the direction perpendicular to the tilting direction and thus trigger a crystallization of parallel stripes. This mechanism is a potential candidate for realizing supersolid state in 2D in the dilute regime. When entering the phonon instability regime, the stripes will dislocate and lose crystalline order~\cite{mishra2016dipolar}.

\subsection{Supersolid in 2D}

Recent advances  have led to the experimental realization of supersolids through the formation of quantum droplets that maintain phase coherence in elongated or planar traps within the 3D regime. The phenomenon is typically understood as a droplet array immersed in a BEC background~\cite{tanzi2019observation,bottcher2019transient,chomaz2019long}. However, these studies have primarily focused on high-dimensions with 3D wave functions. 

Alternatively, one may explore a complementary mechanism for inducing supersolidity in dipolar quantum gases using anisotropy with respect to rotational symmetry in 2D systems with tilted dipoles. This approach provides the additional control parameter of dipole tilt angle, offering new pathways to explore supersolids. However, supersolidity coexisting with a 2D superfluid remains unobserved~\cite{recati2023supersolidity}.


 \begin{figure}[b]
\includegraphics[scale=0.35]{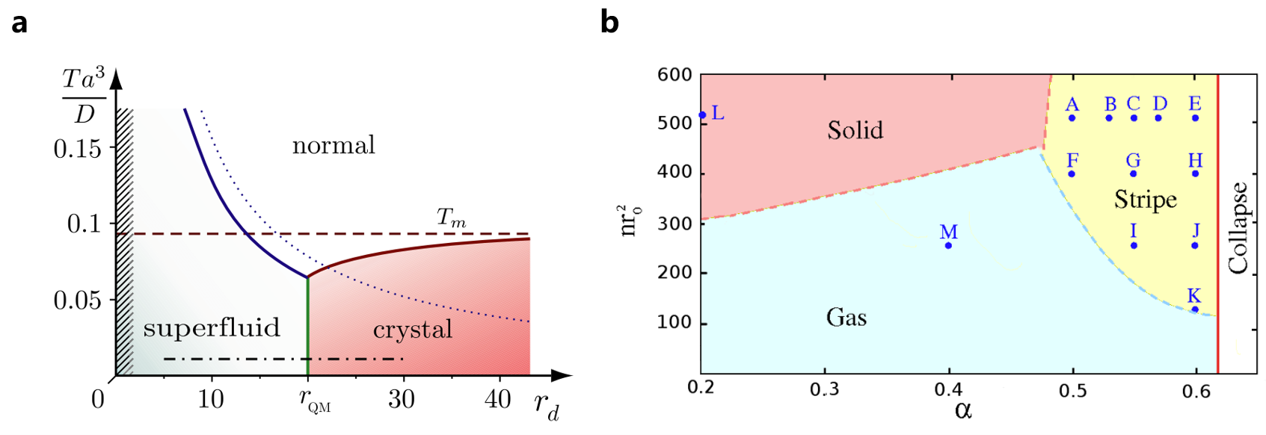}
\caption{\textbf{a.} Phase diagram of 2D dipolar bosons at finite temperature with dipoles perpendicular to the 2D plane. From Ref.~\cite{buchler2007strongly}. \textbf{b.} Zero-temperature phase diagram of 2D dipolar bosons with tilted dipoles. From Ref.~\cite{bombin2017dipolar}.}
\label{crystal}       
\end{figure}

\subsubsection{Strongly interacting dipolar crystal and stripes}
If all dipoles are aligned perpendicular to the 2D plane, they will repel each other and the dipoles can spontaneously organize themselves into a triangular lattice with strong enough DDI~\cite{buchler2007strongly}. The crystal melts with increasing temperature or decreasing dipolar interaction strength, as shown in Fig.~\ref{crystal}a. This strongly correlated quantum phase requires an extremely large dipole moment at low temperature and has not been observed yet. Such a strong dipolar interaction is possible to be achieved in ultracold dipolar molecules~\cite{bigagli2024observation}.

In the strongly interacting regime, when the dipoles are tilted above a critical angle, the ground state of the system is predicted to be a 2D dipolar stripe as shown in Fig.~\ref{crystal}b~\cite{bombin2017dipolar}. One sees that by tilting the dipoles, the phase transition point from gas phase to stripe crystal requires a smaller dipole moment compared with that of solid crystal.

The newly created BEC of dipolar molecules with strong DDI~\cite{bigagli2024observation} is promising to realize such a 2D dipolar crystal. Meanwhile, whether this dipolar crystal in 2D can simultaneously present superfluidity and become a supersolid is still under debate~\cite{bombin2019berezinskii,cinti2019absence}.

\subsubsection{\textbf{Toward supersolids in the quasi-2D regime}} 
 To study 2D physics associated with BKT phenomena in ultracold atom experiments, the quasi-2D conditions $\hbar\omega_z>\mu$ and $\hbar\omega_z>k_BT$ are required~\cite{hadzibabic2008trapped}. In this regime, atoms occupy only the ground state of the harmonic oscillator along $z$, and the quantum-statistic-driven BEC transition is replaced by the interaction-driven BKT transition. However, the harmonic length $l_z$ becomes extremely small, making roton formation difficult. Although supersolid states were predicted in quasi-2D systems with dipoles perpendicular to the 2D plane~\cite{zhang2021phases,ripley2023two}, achieving roton softening to the instability regime typically requires a large atomic density. These conditions are difficult to access in current 2D experiments with magnetic atoms. Consequently, all the experiments investigating supersolid properties with magnetic atoms are completed under the condition that the dynamics of the wave function are 3D.

Recent predictions suggest that in the quasi-2D regime, critical parameters for approaching roton instability—including interaction strength and density—can be significantly reduced by using tilted dipoles and additional short-range interaction cores~\cite{staudinger2023striped,aleksandrova2024density,sanchez2025tilted}. This reduction makes observing 2D supersolids in dilute gases possible. When dipoles are tilted at angle $\theta$ relative to the normal direction, $F(x)$ becomes $(3\cos^2\theta-1)-3\cos^2\theta\sqrt{\pi}xe^{x^2}\rm erfc(\textit{x})$. The intuitive explanation is that when dipoles are tilted beyond the magic angle $\theta\approx54.7^\circ$ (relative to the normal direction), the constant part of the dipole-dipole interaction (DDI) becomes negative. This negative component partially compensates for the repulsive contact interaction, reducing $g_{\text{eff}}$ and making roton instability accessible with finite \textbf{k}. This mechanism highlights the dipolar angle $\theta$ as a crucial tuning parameter for realizing supersolid states in 2D dipolar systems including magnetic atoms~\cite{staudinger2023striped,aleksandrova2024density,sanchez2025tilted}.

\section{Outlook}

As we venture into dipolar quantum systems, numerous exciting possibilities emerge across multiple research directions. The unique properties of dipolar interactions—long-range, anisotropic, and tunable—position these systems as exceptional platforms for exploring fundamental physics as well as simulating novel quantum phenomena. Meanwhile, there are still experimental challenges that need to be overcome in the future studies.

\vspace{10pt}

{\bf Exotic quantum states in 2D}
The 2D dipolar superfluid serves as a versatile platform for exploring exotic quantum states of matter, including phenomena like the anisotropic vortex core and anisotropic vortex-vortex interactions~\cite{mulkerin2013anisotropic}, as well as stable 2D bright solitons~\cite{pedri2005two}. Additionally, it supports the formation of 2D supersolids exhibiting diverse crystalline patterns~\cite{zhang2021phases,ripley2023two,recati2023supersolidity,aleksandrova2024density}. In 2D dipolar superfluids with dipoles tilted toward the plane, theoretical predictions highlight unique anisotropic phase-phase correlations and density-density correlations, offering insights into novel quantum behaviors~\cite{ticknor2012anisotropic,baillie2014number}. Another interesting direction is the study of superfluid turbulence that may exhibit anomalous scaling behavior, particularly near non-thermal fixed points induced by tilted dipoles in a quasi-2D trap~\cite{rasch2025anomalous}. 
\vspace{10pt}

{\bf Cosmological quantum simulation} Tunable atomic systems allow researchers to address various cosmological questions through controlled experiments~\cite{hung2013cosmology,hu2019quantum}, overcoming the traditional challenges of recreating the Universe's initial conditions. Beyond contact interactions in quasi-2D systems, dipolar interactions offer a unique opportunity to simulate analogue gravity phenomena while providing flexibility in controlling dispersion relations~\cite{Chandran.2025}. Trans-Planckian dispersion has been investigated in dipolar 2D systems~
\cite{Cha.2017,Tian.2022} and appears to be within our reach in the near future.

\vspace{10pt}
{\bf Strongly dipolar molecules }
By pairing two magnetic atoms through Feshbach resonance, one can create strongly dipolar molecules with large permanent magnetic moments, such as Er$_2$ formed from bosonic erbium atoms~\cite{frisch2015ultracold}. Similar to ultracold polar molecules, these dipolar molecules benefit from significantly reduced loss rates in 2D geometries, making them promising candidates for observing novel quantum phases including those with unpolarized dipoles.

\vspace{10pt}
{\bf Supersolids in 2D} Exploring supersolids in 2D now appears to be within experimental reach, using both magnetic atoms and polar molecules. By leveraging the anisotropic nature of DDI with tilted dipoles, whether a stripe-type supersolid occurs at specific parameters remains an interesting open question. A fascinating question concerns how a 2D supersolid might simultaneously exhibit both quasi-long-range superfluid order and quasi-crystalline order, spontaneously breaking two symmetries at once, which can be potentially measured in experiment. Another interesting open question is how supersolid states melt in the presence of vortex-dislocation defects. Cold atom platforms may provide a unique opportunity to directly reveal these melting transitions, addressing whether the two separate melting processes—associated with superfluidity and crystallinity—are locked together or independent.

\vspace{10pt}
{\bf Experimental challenges}
To study coherence and excitations in 2D strongly dipolar BKT superfluids, a box potential~\cite{chomaz2015emergence,christodoulou2021observation,chauveau2023superfluid} proves essential for creating the homogeneous conditions needed for thermodynamic study. For example, such a homogeneous system with tunable size allows us to study the scaling behavior of exotic phases like supersolid towards thermodynamic limit, which is crucial to understand the real phase transition. Due to dense Feshbach resonances and broad transitions in blue wavelength of Lanthanide atoms, achieving a far-blue-detuned light is not as straightforward as alkaline atoms. Recently, blue-detuned traps have been demonstrated for dysprosium atoms~\cite{preti2024blue}.

\paragraph{\bf Acknowledgement} This manuscript will appear as a chapter in the Spring book entitled {\it Short and Long
Range Quantum Atomic Platforms — Theoretical and Experimental Developments}, edited by P. G. Kevrekidis, C. L. Hung, and S. I. Mistakidis. G.-B.J acknowledges support by the Gordon and Betty Moore Foundation, grant DOI 10.37807/GBMF13794.

%

\bibliographystyle{unsrt}

\providecommand{\noopsort}[1]{}\providecommand{\singleletter}[1]{#1}%

\end{document}